\documentclass[twocolumn,showpacs,preprintnumbers,amsmath,amssymb,floatfix]{revtex4-1}

\usepackage{graphicx}
\usepackage{amsmath,amsfonts,amssymb}
\usepackage{graphicx}
\usepackage{color}
\usepackage{bbold}

\usepackage[normalem]{ulem}
\usepackage{todonotes}
\usepackage[bookmarks,
            breaklinks,
            pdfstartview=FitH, 
            pdftitle={},
            pdfauthor={},
            pdfkeywords={}
            ]{hyperref}
\usepackage{ulem}
\hypersetup{
  colorlinks   = true,  
  urlcolor     = blue,  
  linkcolor    = blue,  
  citecolor    = red    
}

\usepackage[margin=1.in]{geometry}
\usepackage{graphicx}
\usepackage{amsmath}
\usepackage{amssymb}
\usepackage{mathtools}
\usepackage{amsthm}
\usepackage{tikz-cd}
\usepackage{subfigure}
\usepackage{booktabs}

\begin{document}

\title{Strong spin-exchange recombination of three weakly interacting $^{7}$Li atoms}
\author{J.-L. Li}
\altaffiliation[Corresponding author: ]{j.li1@tue.nl}
\affiliation{Eindhoven University of Technology, P.~O.~Box 513, 5600 MB Eindhoven, The Netherlands}
\author{T. Secker}
\affiliation{Eindhoven University of Technology, P.~O.~Box 513, 5600 MB Eindhoven, The Netherlands}
\author{P. M. A. Mestrom}
\affiliation{Eindhoven University of Technology, P.~O.~Box 513, 5600 MB Eindhoven, The Netherlands}
\author{S. J. J. M. F. Kokkelmans}
\affiliation{Eindhoven University of Technology, P.~O.~Box 513, 5600 MB Eindhoven, The Netherlands}

\begin{abstract}
We reveal a significant spin-exchange pathway in the three-body recombination process for ultracold lithium-7 atoms near a zero-crossing of the two-body scattering length. This newly discovered recombination pathway involves the exchange of spin between all three atoms, which is not included in many theoretical approaches with restricted spin structure. Taking it into account, our calculation is in excellent agreement with experimental observations. To contrast our findings, we predict the recombination rate around a different zero-crossing without strong spin-exchange effects to be two orders of magnitude smaller, which gives a clear advantage to future many-body experiments in this regime. This work opens new avenues to study elementary reaction processes governed by the spin degree of freedom in ultracold gases.
\end{abstract}
\date{\today}
\maketitle
\textit{Introduction}---Weakly interacting ultracold Bose gases are an excellent testbed for fundamental many-body theories due to their theoretical simplicity and experimental controllability and tunability at a high precision level \cite{Andersen:2004,Pethick:2008,Bloch:2008}. In most of the experimental realizations \cite{Bloch:2008,Weber:2003,Eismann:2016,Donley:2001,Altin:2011,Schemmer:2018,Dogra:2019}, three-body recombination (TBR) is very crucial as it constitutes a major loss source and usually determines the lifetime of the ultracold cloud.  In addition, TBR has also been demonstrated to cause anti-evaporative heating \cite{Weber:2003,Eismann:2016}, interplay with collapse dynamics \cite{Donley:2001,Altin:2011} and is predicted to cool and even purify ultracold ensembles under particular conditions \cite{Schemmer:2018,Dogra:2019}.  So far, there have been only a few experimental investigations on TBR rates in the weak interaction regime, particularly concerning the magnetic field dependence of the TBR \cite{Kraemer:2006,Shotan:2014}. 
Theoretically, quantifying TBR rates in this context is challenging and a numerical approach for it remains highly desirable.

TBR occurring in ultracold atomic gases is also a good candidate for understanding fundamental chemistry given that the reactants can be prepared in a full quantum regime with extremely high control over all external and internal degrees of freedom \cite{Balakrishnan:2016,Quemener:2012,Greene:2017}.  The entire reaction process has been well understood for strongly repulsive ultracold atoms for which the recombination into the shallowest molecular product is prominent \cite{braaten:2006,Kraemer:2006,Pollack:2009,Gross:2009,Gross:2010,Zaccanti:2009}. In the weak interaction regime, however, the reaction pathways for TBR  can be much more complicated and the investigation of product distributions is very challenging. Early studies on this subject were established using the Jastrow approximation \cite{Moerdijk:1996} or limited to simple systems for which only a few molecular products are involved \cite{Wang:2011,Suno:2009}. 

Inspiringly, experimental milestones have been achieved in the past few years by combining hybrid atom-ion traps and resonance-enhanced multiphoton ionization techniques \cite{Harter:2013,Wolf:2017,Wolf:2019}. In such experiments, chemical reaction pathways are identified on the level of full quantum state-to-state resolution regarding the electronic, vibrational, rotational, hyperfine and even magnetic quantum numbers. Several propensities in TBR processes, such as two atoms conserving their total parity,  total spin and magnetic projection of total spin when forming weakly bound molecular products, are established for $^{87}$Rb atoms and also explained using the hypothesis that the third atom does not flip its internal spin. Although frequently implemented in previous works for enabling three-body calculations \cite{Wang:2014,Kato:2017,Chapurin:2019,Xie:2020}, this hypothesis may not be generally valid, as is indicated in Ref. \cite{Secker:2021} for strongly interacting $^{39}$K atoms. Therefore, whether the hypothesis and the propensities established in the $^{87}$Rb system are applicable for other species remains an open question.

In this Letter, we investigate the TBR process for weakly interacting ultracold $^{7}$Li atoms in an external magnetic field using a multichannel framework. We successfully quantify the TBR rate and identify the dominant recombination pathways. One of these pathways involves spin-exchange between the created molecule and the remaining free atom. This demonstrates the violation of the aforementioned hypothesis for $^{7}$Li and we further analyze the origin of this violation. For comparison, we also decrease the magnetic field to study the TBR process of $^{7}$Li atoms in the regime where spin-exchange decay processes are much less significant.

\textit{Spin models}---We follow Ref. \cite{Secker:2021}  to write down the Hamiltonian $H_{0}$, describing three alkali-metal atoms at infinite separation in an external magnetic field $B$, as
\begin{equation}
H_0 = \sum_{\sigma_1 \sigma_2 \sigma_3} \left(T + E_{\sigma_1} + E_{\sigma_2} + E_{\sigma_3} \right) | \sigma_1 \sigma_2 \sigma_3 \rangle \langle \sigma_1 \sigma_2 \sigma_3 | \label{H0} ,
\end{equation}
where $T$ is the kinetic energy operator, and $E_{\sigma_{a}}$ and $|\sigma_{a}\rangle$ denote the channel energy and the corresponding internal spin state, respectively, of atom $a$ $(a=1,2,3)$, which are $B$-dependent. Adiabatically, each state $|\sigma\rangle$ can be unambiguously traced back to a hyperfine state $|f,m_f\rangle$ at zero field, or forward to a $|m_s,m_i\rangle$ state at infinite field \cite{Chin:2010}. Here, $f$ denotes the quantum number of atomic total spin $\mathbf{f}$ summing up the electronic spin $\mathbf{s}$ and nuclear spin $\mathbf{i}$, and $m_f$, $m_i$ and $m_s$ are the corresponding magnetic quantum numbers. Even though commonly $(f,m_f)$ is used for labeling $\sigma$, we note that $(m_s,m_i)=\sigma$ is more appropriate in this work given that the considered magnetic fields are high. 

In addition to $H_0$,  we assume that the three atoms interact pairwisely, $V=V_{12}+V_{23}+V_{13}$. The pairwise potential
\begin{equation} \label{VIJ}
 V_{ab}=V_{ab}^S (r_{ab}) \mathcal{P}_{ab}^S + V_{ab}^T (r_{ab}) \mathcal{P}_{ab}^T,
 \end{equation}
  consists of singlet $V_{ab}^{S}$ and triplet $V_{ab}^{T}$ components
in the electronic ground configuration of two alkali-metal atoms, where $r_{ab}$ represents the distance between atom $a$ and $b$ and $\mathcal{P}_{ab}^{S}$ $(\mathcal{P}_{ab}^{T})$ describes the projector on the electronic singlet (triplet) state of pair ($a,b$). We use realistic molecular potentials for $V_{ab}^{S}$ and $V_{ab}^{T}$ in this work. We refer to the interaction model in Eq. (\ref{VIJ}) as the full multichannel spin (FMS) model.
Several approximations can be made for simplifying the three-body calculation by restricting the way the atoms interact with each other.  One frequently used restriction requires the third (spectating) atom to be fixed to the initial spin state for the other two atoms to interact, referred to as the fixed spectating spin (FSS) model in Ref. \cite{Secker:2021}. The pairwise interaction under such restriction is expressed as
\begin{equation} \label{VIJS}
V_{ab}^{\rm{FSS}} = V_{ab} | \sigma^{\rm{in}}_c\rangle \langle \sigma^{\rm{in}}_c |, 
\end{equation}
where $| \sigma^{\rm{in}}_c\rangle$ denotes the initial spin state of atom $c$ and $(a,b,c)=(1,2,3), (2,3,1)$ or $(3,1,2)$. We also construct an optimized spin (OPS) model via
\begin{equation} \label{VIJO}
V_{ab}^{\rm{OPS}} = \sum_{\sigma_c \in \mathcal{D}_{c}} V_{ab} | \sigma_c\rangle \langle \sigma_c |,
\end{equation}
where $\mathcal{D}_c$ represents the spin states of atom $c$ that play a dominant role in the collision. It is apparent that the OPS model is equivalent to the FSS model when $\mathcal{D}_{c} \rightarrow \{ \sigma^{\rm{in}}_c \}$ and to the FMS model when $\mathcal{D}_c$ includes all single-particle spin states.

Once the spin model is given, we use the Alt-Grassberger-Sandhas (AGS) equation \cite{Alt:1967} to calculate the partial TBR rate $K_3^d$ for the decay process into a specific atom-molecule channel at zero collisional energy, see supplemental material \cite{sm}. Here, $d$ labels both the molecule and the corresponding decay channel. The total rate,
 $K_3=\sum_{d}K_3^{d}$, therefore sums up all partial contributions. We define $K_3$ such that $dn/dt=-\frac{1}{2}K_3 n^3$,
where $n$ denotes the density of the atomic gas. This definition is consistent with the one in Refs. \cite{Braaten:2008,Mestrom:2019,Secker:2021,Secker:2021map}, while it differs from the one in Ref. \cite{Shotan:2014} by a factor of two. In our calculations, we truncate the molecular orbital angular momentum quantum number $l$ up to $l_{\rm{max}}$ and implement a cut-off on the relative momentum between the atom and the molecule \cite{sm}.

\textit{Strong spin-exchange TBR}---We investigate a system of three $^{7}$Li atoms at zero energy initially prepared in the same $|m_{s}=-1/2,m_{i}=1/2\rangle$ state, which corresponds to $|f=1,m_f=0 \rangle$ in conventional notation. We study the system in an external magnetic field varied between 847 and 885 G, covering a zero-crossing of the two-body $s$-wave scattering length at $B = 850$ G, where experimental data for the TBR rate are reported in Ref. \cite{Shotan:2014}. The singlet and triplet potentials are taken from Ref. \cite{Julienne:2014}. For comparison, we follow  Ref. \cite{Shotan:2014} to define the recombination length $L_m$ via
\begin{equation}
K_3=328.2\frac{\hbar}{m}L_m^4,
\end{equation}
where $m$ denotes the mass of the atom.

\begin{figure}[t]
\centering
 \resizebox{0.5\textwidth}{!}{\includegraphics{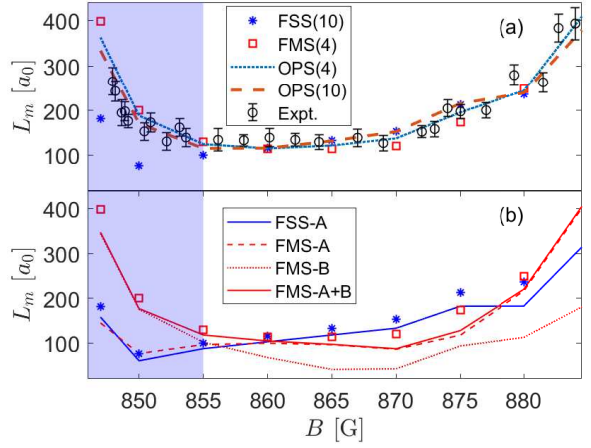} }
\caption{\label{fig:li10} (a) There-body recombination length in units of the bohr radius $a_0$ for $^{7}$Li in the $|m_s=-1/2,m_i=1/2\rangle$ state. The stars denote the result from the FSS model with $l_{\rm{max}}=10$ while the squares denote the results of the FMS model with $l_{\rm{max}}=4$. The dotted and dashed lines correspond to the calculations using the OPS model with $l_{\rm{max}}=4$ and 10, respectively. The experimental data at 2.5 $\mu$K are taken from Ref. \cite{Shotan:2014}.  (b) The corresponding partial contributions of the decay channels A and B in the FSS($l_{\rm{max}}=10$) and FMS($l_{\rm{max}}=4$) calculations. The stars and squares represent the same results as in (a). The light purple area indicates the strong three-body spin-exchange regime.}
\end{figure}

Figure \ref{fig:li10}(a)  compares our results to the experimental measurement in Ref. \cite{Shotan:2014}.  The result of the FMS calculation with $l_{\rm{max}}=4$ is in excellent agreement with the experimental results at the considered magnetic fields.  However, that of the FSS calculation with $l_{\rm{max}}=10$ only agrees with experiment for $B\gtrsim$ 860 G, but deviates from the experimental measurement for $B\lesssim$ 860 G. The difference in performance between the two approaches does not result from the truncation of the quantum number $l$, since our additional FSS calculation with $l_{\rm{max}}=4$ leads to only a small shift compared to  that with $l_{\rm{max}}=10$, not shown in Fig. \ref{fig:li10}(a) though. Therefore, we attribute the invalidity of the FSS model to its incapability to represent some important three-body channels.

By analyzing our FMS result, we find two dominant product channels, which together contribute more than 50$\%$ to the total TBR rate among more than 300 involved products in our model. We identify that one dominant product channel (decay channel A) consists of the energetically shallowest $s$-wave molecule with a projection quantum number of total two-body spin $M_{2b}=m_{s_a}+m_{i_a}+m_{s_b}+m_{i_b}=0$ plus a free atom in its initial $|m_{s_c}=-1/2,m_{i_c}=1/2\rangle$ spin state. This decay channel is included in the FSS model and the corresponding contributions to the recombination length are similar in both the FSS and FMS calculations as is shown in Fig. \ref{fig:li10}(b).  However, the other dominant product channel (decay channel B) consisting of the shallowest $s$-wave molecule with $M_{2b}=-1$ plus a free atom in the $|m_{s_c}=-1/2,m_{i_c}=3/2\rangle$ state is not represented in the FSS model. The spin-exchange recombination to decay channel B becomes increasingly important with decreasing magnetic field at $B\lesssim$ 860 G and ultimately dominates over the one to decay channel A when $B\lesssim$ 855 G, leading to a rapid enhancement of the loss rate. Moreover, its contribution matches very well with the deviation between the FSS calculation and the measurement at $B\lesssim$ 860 G.

 Allowing the third atom to switch from its initial  $|m_{s_c}=-1/2,m_{i_c}=1/2\rangle$ to the $|m_{s_c}=-1/2,m_{i_c}=3/2\rangle$ spin state, we arrive at an OPS model with $\mathcal{D}_{c}=\{(-1/2,1/2),(-1/2,3/2)\}$. This OPS model gives results consistent with the measurement, when we truncate the molecular orbital angular momentum quantum number $l$ at $l_{\rm{max}}=4$ and 10. The results are also in line with the values we calculated with the FMS model, demonstrating again that the truncation of $l$ has a minor influence on our calculation. 

 In Ref. \cite{Shotan:2014}, the enhancement of the TBR rate at $B\lesssim$ 855 G is suggested to be governed by a two-body length scale $L'_e$ that is determined by the two-body scattering length and effective range parameter. In contrast, our analysis demonstrates that it originates from a single specific atom-molecule product channel coupled to the three-body incoming state via the spin-exchange recombination mechanism. In general, the two-body quantity $L'_e$ is not able to describe this three-body spin-exchange process.  It remains an open question why $L'_e$ works beyond its capacity to explain the TBR rate qualitatively \cite{Shotan:2014}.

The three-body channels with $|m_{s_c}=-1/2,m_{i_c}=3/2\rangle$ become important in the present calculation because their small energy separations to the incoming threshold lead to large multichannel couplings \cite{Secker:2021,sm}. However, the observed effect that three $^{7}$Li atoms recombine predominantly into decay channel B seems counterintuitive at first glance since decay channel A is less separated from the three-body incoming threshold than decay channel B in the considered magnetic field regime \cite{sm}. To explain the strong recombination into decay channel B, we use an approach similar to the one in Ref. \cite{Wolf:2017}, in which the TBR rate to a specific decay channel is explained by the overlap of the product molecule state and a zero energy scattering state of two atoms forming that molecule. However, the original treatment of Ref. \cite{Wolf:2017} is based on the hypothesis that the third atom does not flip its internal spin during the TBR process and cannot describe the recombination process into decay channel B. Therefore we extend this treatment by taking into account the interaction with the third atom and the exact three-body spin structure and study the overlap $\mathcal{O}_d= {}_\alpha\langle\psi_d|(P_++P_-)V_\alpha|\psi_{\rm{scat}}\rangle_\alpha$ \cite{sm}. We use $\alpha=(a,b)$ to label the pair $(a,b)$ that forms the molecule $d$. $|\psi_d\rangle_\alpha$ and $|\psi_{\rm{scat}}\rangle_\alpha$ denote the state of molecule $d$ plus a free atom and that of a zero-energy scattering complex of the pair $(a,b)$  plus a free atom, respectively \cite{fn}. The interaction term $(P_++P_-)V_\alpha$ that couples  $|\psi_{\rm{scat}}\rangle_\alpha$ and $|\psi_d\rangle_\alpha$ is derived from a perturbative analysis on the AGS equation \cite{sm}.  $P_+$ and $P_-$ denote the cyclic and anticyclic permutation operators, respectively. We use $\mathcal{O}_d$ to explain the dominancies of decay channels A and B as it captures the overall trend and relative magnitude of the TBR rates at the considered magnetic fields \cite{sm}.

\begin{figure}[t]
\centering
 \resizebox{0.5\textwidth}{!}{\includegraphics{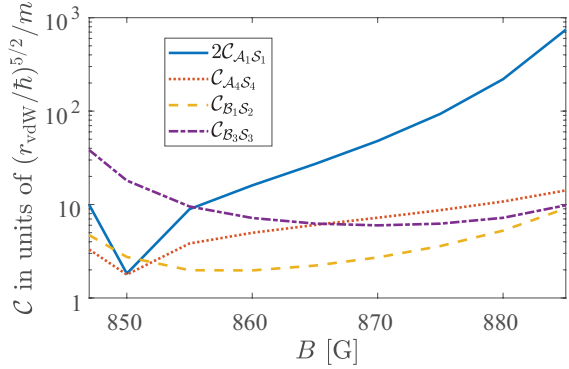} }
\caption{\label{fig:ab} Components of $\mathcal{O}_A$ and $\mathcal{O}_B$ as a function of magnetic field. Here $r_{\rm{vdW}}$ denotes the characteristic length scale for the van der Waals interaction between two atom \cite{Chin:2010}.}
\end{figure}

  \begin{table*}
  \tabcolsep=6pt
\small
\renewcommand\arraystretch{1.0}
\caption{\label{tab:sd} Nonzero elements of $\langle\sigma_{2b}^d\sigma_c^d| P_+^{\rm{s}}|\sigma_{2b}^{\rm{scat}}\sigma_c^{\rm{in}}\rangle$ and the corresponding spin states for decay channels A and  B. We use $\sigma_{2b}^A=\{\mathcal{A}_1,\mathcal{A}_2,\cdots,\mathcal{A}_8\}$, $\sigma_{2b}^B=\{\mathcal{B}_1,\mathcal{B}_2,\cdots,\mathcal{B}_6\}$ and $\sigma_{2b}^{\rm{scat}}=\{\mathcal{S}_1,\mathcal{S}_2,\cdots,\mathcal{S}_8\}$ to label these two-body spin states in the order of increasing two-body channel energy. Note that $\sigma_c^{\rm{in}}=(-1/2,1/2)$ in all cases.}
 \begin{tabular}{cccccc}
 \hline
 \hline
$d$&$\sigma_c^{d}$&$\sigma_{2b}^{d}$&$\sigma^{\rm{scat}}_{2b}$&$\langle\sigma_{2b}^d\sigma_c^d| P_+^{\rm{s}}|\sigma_{2b}^{\rm{scat}}\sigma_c^{\rm{in}}\rangle$ \\ 
\hline
A&$(-1/2,1/2)$&$\mathcal{A}_1=(-1/2,1/2;-1/2,1/2)$&$\mathcal{S}_{1} =(-1/2,1/2;-1/2,1/2)$&1  \\ 
A&$(-1/2,1/2)$&$\mathcal{A}_4 = (-1/2,1/2;1/2,-1/2)$&$\mathcal{S}_{4}=(-1/2,1/2;1/2,-1/2)$&1/2 \\ 
B&$(-1/2,3/2)$&$\mathcal{B}_1 =(-1/2,-1/2;-1/2,1/2)$&$\mathcal{S}_{2}=(-1/2,-1/2;-1/2,3/2)$&1/2 \\ 
B&$(-1/2,3/2)$&$\mathcal{B}_3=(1/2,-3/2;-1/2,1/2)$&$\mathcal{S}_{3}= (1/2,-3/2;-1/2,3/2)$&1/2 \\ 
 \hline
 \hline
 \end{tabular}
 \end{table*}

The overlaps $\mathcal{O}_A$ and $\mathcal{O}_B$ are written as 
\begin{equation}
\mathcal{O}_d=2\sum_{\sigma_{2b}^d,\sigma_{2b}^{\rm{scat}}} \mathcal{C}_{\sigma_{2b}^d\sigma_{2b}^{\rm{scat}}}\langle\sigma_{2b}^d\sigma_c^d| P_+^{\rm{s}}|\sigma_{2b}^{\rm{scat}}\sigma_c^{\rm{in}}\rangle, d=A,B, \label{OD}
\end{equation}
where $ \mathcal{C}_{\sigma_{2b}^d\sigma_{2b}^{\rm{scat}}}\equiv  {}_\alpha\langle\psi_d|P_+^{\rm{c}}|\sigma_{2b}^{d}\sigma_c^{d}\rangle\langle\sigma_{2b}^{\rm{scat}}\sigma_c^{\rm{in}}|V_\alpha|\psi_{\rm{scat}}\rangle_\alpha$ represents the spatial part of $\mathcal{O}_d$, which can be simplified as $ \mathcal{C}_{\sigma_{2b}^d\sigma_{2b}^{\rm{scat}}}=\langle\phi_d|\frac{1}{2}q_d,\sigma_{2b}^d\rangle\langle q_d,\sigma_{2b}^{\rm{scat}} |V_\alpha|\phi_{\rm{scat}}\rangle_\alpha$  \cite{sm}. We use $P_+^{\rm{s}}$ and $P_+^{\rm{c}}$ to denote the permutation operator $P_+$ acting only on the spin and coordinate space, and $\sigma_{2b}=(m_{s_a},m_{i_a} ;m_{s_b},m_{i_b})$ for the spin state of the pair ($a,b$). We assume that $|\sigma_{2b}\rangle$ is properly symmetrized as in Ref. \cite{Secker:2021}. Here $\phi_{d}$ and $\phi_{\rm{scat}}$ represent the radial wave functions of molecule $d$ and the two-body scattering state, respectively \cite{sm}. Furthermore, $q_d$ denotes the magnitude of the relative momentum between the free atom and molecule $d$.

We find that the field-independent spin coupling matrix $\langle\sigma_{2b}^d\sigma_c^d| P_+^{\rm{s}}|\sigma_{2b}^{\rm{scat}}\sigma_c^{\rm{in}}\rangle$ picks out only a few specific elements of $\mathcal{C}$ that contribute to the overlap $\mathcal{O}_d$. For both decay channels A and B, there are only two contributions of which the corresponding spin states and spin coupling matrix elements are listed in Table \ref{tab:sd}.  
Implementing the results of Table \ref{tab:sd} into Eq. (\ref{OD}), we get
\begin{eqnarray} \label{eq:OAB}
 \mathcal{O}_A&=&2\mathcal{C}_{\mathcal{A}_1\mathcal{S}_1} +\mathcal{C}_{\mathcal{A}_4\mathcal{S}_4},\nonumber\\
 \mathcal{O}_B&=&\mathcal{C}_{\mathcal{B}_1\mathcal{S}_2}+\mathcal{C}_{\mathcal{B}_3\mathcal{S}_3}.
\end{eqnarray}
The above expression shows explicitly that the interplay between two specific elements of the spatial part matrix $\mathcal{C}$ determines the overlap $\mathcal{O}_d$. Figure \ref{fig:ab} shows that $2\mathcal{C}_{\mathcal{A}_1\mathcal{S}_1}$ and $\mathcal{C}_{\mathcal{B}_3\mathcal{S}_3}$ are the most significant contributions at $B \gtrsim 855$ G and $B \lesssim 855$ G, respectively. The enhanced behavior of $2\mathcal{C}_{\mathcal{A}_1\mathcal{S}_1}$ and $\mathcal{C}_{\mathcal{B}_3\mathcal{S}_3}$ thus explains the dominancy of decay channels A and B in each magnetic field regime. We find that these enhancements originate from the influence of the Feshbach resonance at $B = 894$ G on the molecular wave function $\phi_A$ and that of the Feshbach resonance at $B = 845$ G on the two-body scattering wave function $\phi_{\rm{scat}}$, respectively \cite{sm}.

\begin{table}
 \tabcolsep=2pt
\small
\renewcommand\arraystretch{1.0}
 \caption{\label{tab:k3} $K_3$ from the FMS and FSS models for $^{7}$Li at $B = 850$ G (H) and $B = 578$ G (L), and for $^{87}$Rb in the $|f=1,m_f=-1\rangle$ state at $B = 1$ G. In these calculations, we take $l_{\rm{max}}=4$ for $^{7}$Li and $l_{\rm{max}}=10$ for $^{87}$Rb. The numbers are presented in units of cm$^{6}$/s. The singlet and triplet potentials for $^{87}$Rb are taken from Ref. \cite{Strauss:2010}.} 
 \begin{tabular}{cccc}
 \hline
 \hline
Atom&FSS&FMS&Expt. \\
\hline
$^{7}$Li(H)&$1.0\times 10^{-27}$&$3.8\times 10^{-26}$&$1.3(0.4)\times 10^{-26}$ \cite{Shotan:2014}\\
$^{7}$Li(L)&$1.7\times 10^{-28}$&$1.2\times 10^{-28}$ &$<2.3\times 10^{-27}$\cite{Shotan:2014}\\
$^{87}$Rb&$4.0\times 10^{-29}$&$4.0\times 10^{-29}$&$8.6(3.6)\times 10^{-29}$\cite{Burt:1997}  \\
 \hline
 \hline
 \end{tabular}
 \end{table} 
 
\textit{Lower field zero-crossing and comparison to $^{87}$Rb}---For comparison, we investigate the TBR rate near a different zero-crossing of the two-body scattering length at $B = 578$ G in the same spin state of $^{7}$Li \cite{fn2}.
 Table \ref{tab:k3} shows that $K_3$ at $B = 850$ G is higher by two orders of magnitude than that at $B = 578$ G, where the comparable values of $K_3$ predicted by the FSS and FMS models indicate no strong spin-exchange process. We note that decay channel B becomes less important at  $B = 578$ G \cite{sm}. However, the scenario of $^{7}$Li at $B = 578$ G is still in contrast with the $^{87}$Rb system, where the FSS and FMS calculations yield nearly identical results. Therefore, we conclude that the model with the fixed spectating atom's spin state works very well for $^{87}$Rb at low magnetic fields, but not for $^{7}$Li. In general, the TBR rates from our calculation agree with experimental values \cite{Shotan:2014,Burt:1997} within a factor of 2 or 3. The deviation could come from our numerical truncations or from the experimental uncertainty in the number of atoms. For instance, by implementing a larger $l_{\rm{max}}=10$ with the OPS model we get $K_3=2.0\times 10^{-26}$ cm$^{6}$/s at $B =850$ G, which agrees better with the experimentally measured value $1.3(0.4)\times 10^{-26}$ cm$^{6}$/s \cite{Shotan:2014}. 
 
 Our prediction of low TBR rate suggests that $^{7}$Li at $B \approx 578$ G is a good candidate for the first realization of quantum gas purification experiments via three-body loss \cite{Dogra:2019} and the creation of big time crystals \cite{Giergiel:2020}. Other interesting phenomena like the matter wave bright soliton \cite{Khaykovich:2002,Strecker:2002,Nguyen:2014,Luo:2020} and the weak collapse of a Bose-Einstein condensate \cite{Eigen:2016}, which have been experimentally studied, can also be investigated in this specific case, where an extremely small slope $0.01 a_0$/G of the two-body scattering length to magnetic field leads to easy and precise control of the required weak attractive interaction.

\textit{Conclusion}--- We have studied the three-body recombination process of ultracold $^{7}$Li atoms near two zero-crossings of the two-body scattering length at $B =$ 850 G and 578 G. In the vicinity of 850 G, we get a very good agreement with the measured recombination rate and reveal a prominent spin-exchange three-body recombination pathway requiring one atom to flip its nuclear spin when the other two colliding atoms form a molecule. We attribute the prominence of this pathway to the influence of the Feshbach resonance at $B=$ 845 G on the two-body scattering wave function. The strong spin-exchange effect increases the recombination rate by about two orders of magnitude compared to our results around 578 G in the same spin state. Our approach can also be applied to other species to explore the complicated but important multichannel three-body recombination process and to analyze the rich interplay between the translational, vibrational, rotational, electronic spin and nuclear spin degrees of freedom. 

\textit{Acknowledgements}---We thank Lev Khaykovich, Denise Ahmed-Braun, Victor Colussi, Gijs Groeneveld, and Silvia Musolino for discussions.
This research is financially supported by the
Netherlands Organisation for Scientific Research (NWO)
under Grant No. 680-47-623.

\bibliography{biblio}

\begin{thebibliography}{50}
\bibitem [{fn()}]{fn}%
  \BibitemOpen
  \href@noop {} {}\bibinfo {note} {One should not confuse
  $|\psi_{\rm{scat}}\rangle_\alpha$ with the system's real initial state
  $|\psi_{\rm{in}}\rangle$ decribing three free atoms.}\BibitemShut {Stop}%
\end{thebibliography}
\clearpage

\onecolumngrid
\begin{center}
  \textbf{\large Supplemental Material: ``Strong spin-exchange recombination of three weakly interacting $^{7}$Li atoms''}\\[.2cm]
J.-L. Li, T. Secker, P. M. A. Mestrom, and S. J. J. M. F. Kokkelmans\\[.1cm]
  {\itshape Eindhoven University of Technology, P.~O.~Box 513, 5600 MB Eindhoven, The Netherlands\\}
(Dated: \today)\\[1cm]
\end{center}
\twocolumngrid

\setcounter{equation}{0}
\setcounter{figure}{0}
\setcounter{page}{1}
\renewcommand{\theequation}{S\arabic{equation}}
\renewcommand{\thefigure}{S\arabic{figure}}
\renewcommand{\thetable}{S\arabic{table}}  
\renewcommand{\bibnumfmt}[1]{[S#1]}
\section{AGS equation and TBR rate}
We solve the AGS equation in momentum space \cite{Secker:2021,Secker:2021map,Lee:2007,Mestrom:2019}
\begin{align} \label{eq:AGSeqrecom}
U_{\alpha 0} (z) & = \frac{1}{3}G_0^{-1}(z) \left[1 + P_+ + P_-\right] \nonumber \\
& \phantom{=} + \left[ P_+ + P_- \right] \mathcal{T}_\alpha (z) G_0(z) U_{\alpha 0} (z) \, .
\end{align}
via a numerical approach combining the separable expansion method and the two-body mapped grid technique \cite{Secker:2021,Secker:2021map}. Here $G_0=(z-H_0)^{-1}$ is the free Green's operator and $\mathcal{T}_\alpha=(1 - V_\alpha G_0 (z))^{-1} V_\alpha$ represents the generalized two-body transition operator for the pair $\alpha=(a,b)$. $P_+$ and $P_-$ denote the cyclic and anticyclic permutation operators, respectively. The three-body transition operator $U_{\alpha0}$, whose elements describe the transition probabilities from the initial free-atom state to product states of a molecule plus a free atom, is closely related to the TBR rate $K_3$.  In this paper, we define the partial recombination rate $K_3^{d}$ to each specific molecular product $d$ as \cite{Secker:2021,Secker:2021map,Lee:2007,Mestrom:2019}
\begin{equation} \label{eq:K3}
K_3^{d}= \frac{24 \pi m}{ \hbar} (2 \pi \hbar)^6 q_d |{}_\alpha \langle\psi_d | U_{\alpha 0} (z) | \psi_{\rm{in}} \rangle |^2, 
\end{equation}
where $| \psi_{\rm{in}}\rangle$ and $|\psi_d\rangle$ represent the initial and product states, respectively. $q_d$ is the magnitude of the momentum of the free atom relative to the center of mass of molecule $d$. 
In our calculations, we take the zero energy limit $z\rightarrow 0$ from the upper half of the complex energy plane and therefore fix the total orbital angular momentum quantum number $J=0$. The projection quantum number of the total spin angular momentum $M_{\rm{tot}}=\sum_a m_{s_a}+\sum_a m_{i_a}$ ($a=1,2$ and 3) should also be fixed during the scattering process. We also implement truncations  $l_{\max}$ on the orbital angular momentum quantum number $l$ related to the relative movement of the atoms constituting the molecule and $q_{\rm{max}}$ on the magnitude of the momentum $q$ of the third atom relative to the molecule's center of mass. In particular, $q_{\rm{max}}= 20$ $\hbar/r_{\rm{vdW}}$ is used throughout the entire paper and $l_{\max}$ is stated explicitly when the results are presented in the main text. It is worth noting that the sufficiency of $q_{\rm{max}}= 20$ $\hbar/r_{\rm{vdW}}$ is demonstrated for addressing the three-body parameter in Refs. \cite{Secker:2021,Secker:2021map} and additionally checked for our present study by comparing to a larger cutoff $q_{\rm{max}}= 40$ $\hbar/r_{\rm{vdW}}$.  For more details about our numerical approach, we refer the reader to Refs. \cite{Secker:2021,Secker:2021map}. 
\section{Three-body channel energy}
  \begin{figure}[t]
\centering
 \resizebox{0.45\textwidth}{!}{\includegraphics{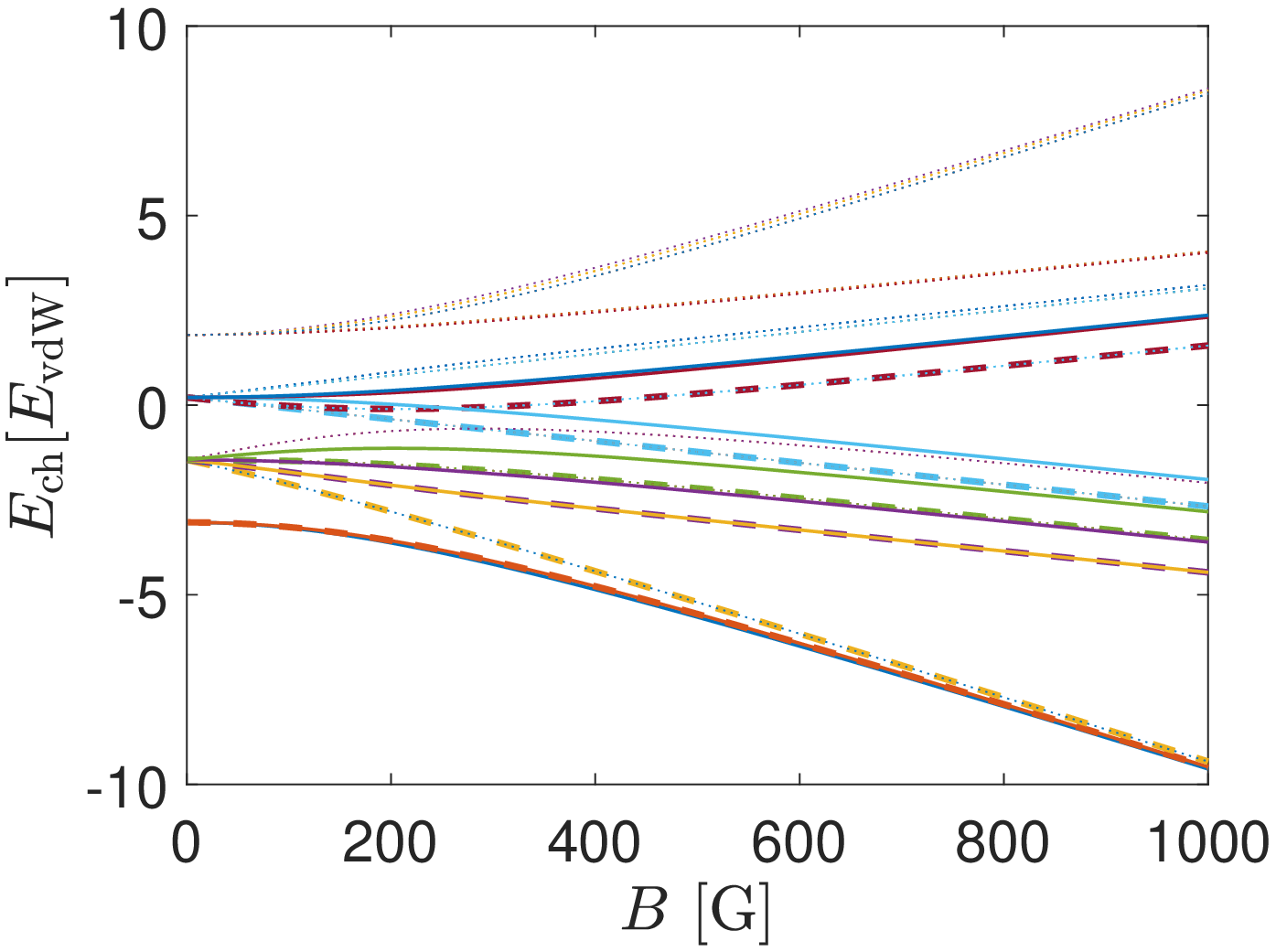} }
\caption{\label{fig:zm} Channel energy $E_{\rm{ch}}$ in units of $E_{\rm{vdW}}=\hbar^2/(mr^2_{\rm{vdW}})$ of three $^{7}$Li atoms with $M_{\rm{tot}} =0$. The solid and dashed lines represent the three-body channels with  $\sigma_c$ = $(-1/2,1/2)$ and $(-1/2,3/2)$, respectively. The dotted lines correspond to other $\sigma_c$. The incoming three-body channel is the lowest one.}
\end{figure}
Figure \ref{fig:zm} shows the three-body channel energy $E_{\rm{ch}} = E_a+E_b+E_c$ with $M_{\rm{tot}}=0$ for $^{7}$Li atoms. One can see that the channel energy separations are in general smaller than those for $^{39}$K atoms \cite{Secker:2021}. In particular for those with $\sigma_c=(-1/2,3/2)$, the two lowest channels are extremely close, with energy separations less than 0.25 $E_{\rm{vdW}}$, to the three-body incoming channel when $B > 800$ G. According to the analysis in Ref. \cite{Secker:2021}, this can lead to strong multichannel couplings to the incoming channel.

\section{Asymptotic energy of decay channels A and B}
  \begin{figure}[t]
\centering
 \resizebox{0.5\textwidth}{!}{\includegraphics{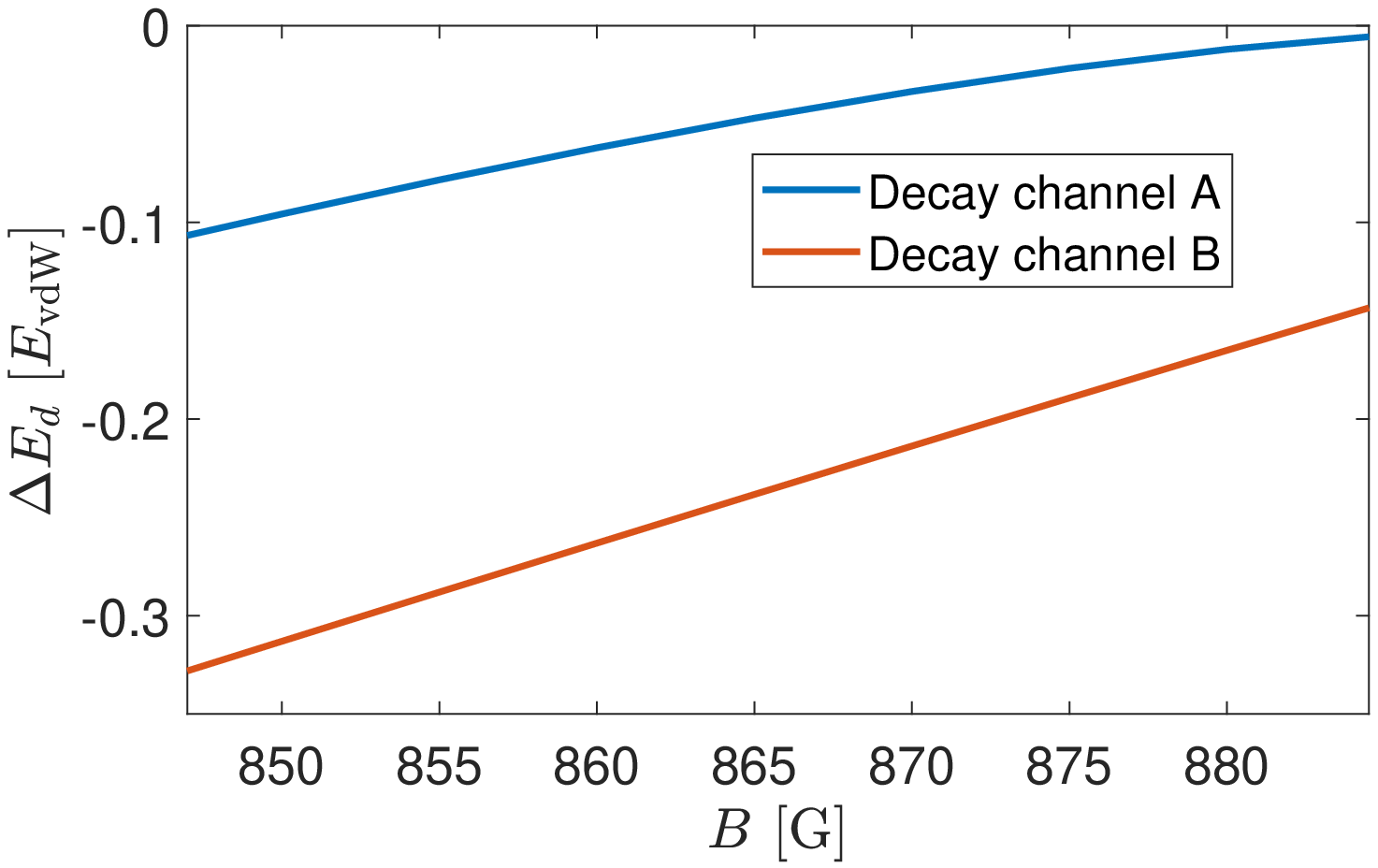} }
\caption{\label{fig:Eb} Energy sperations $\Delta E_d$ of decay channels A and B from the incoming threshold as a function of magnetic field.}
\end{figure}
To illustrate the energy separations of the product channels from the incoming channel, we calculate $\Delta E_d=E_{2b}^{d}+E_{\sigma_{c}^{d}}-E_{\sigma_a^{\rm{in}}}-E_{\sigma_b^{\rm{in}}}-E_{\sigma_c^{\rm{in}}}$ for $d= A$ or $B$, where $E_{2b}^{d}$ denotes the energy level of molecule $d$ and $E_{\sigma_{c}^{d}}$ represents the corresponding shift due to the third  atom. In the zero energy limit considered in this work,  $\Delta E_d$ is simply connected to $q_d$ via $\Delta E_d=-3q_d^2/4m$. Figure \ref{fig:Eb} shows the energy levels of both decay channels are shifted towards the incoming threshold with the increase of the magnetic field and $|\Delta E_{A}|$ persists to be smaller than  
$|\Delta E_{B}|$ in the considered magnetic field regime. These energy separations explain the dominancy of decay channel A at $B\gtrsim$ 860 G. However, it is in contrast with our observation that  $K_3^{B}$ is much larger than $K_3^{A}$ at $B\lesssim$ 855 G. 

\section{TBR rate from $\mathcal{O}_d$} 
    \begin{figure*}[t]
\centering
 \resizebox{1.0\textwidth}{!}{\includegraphics{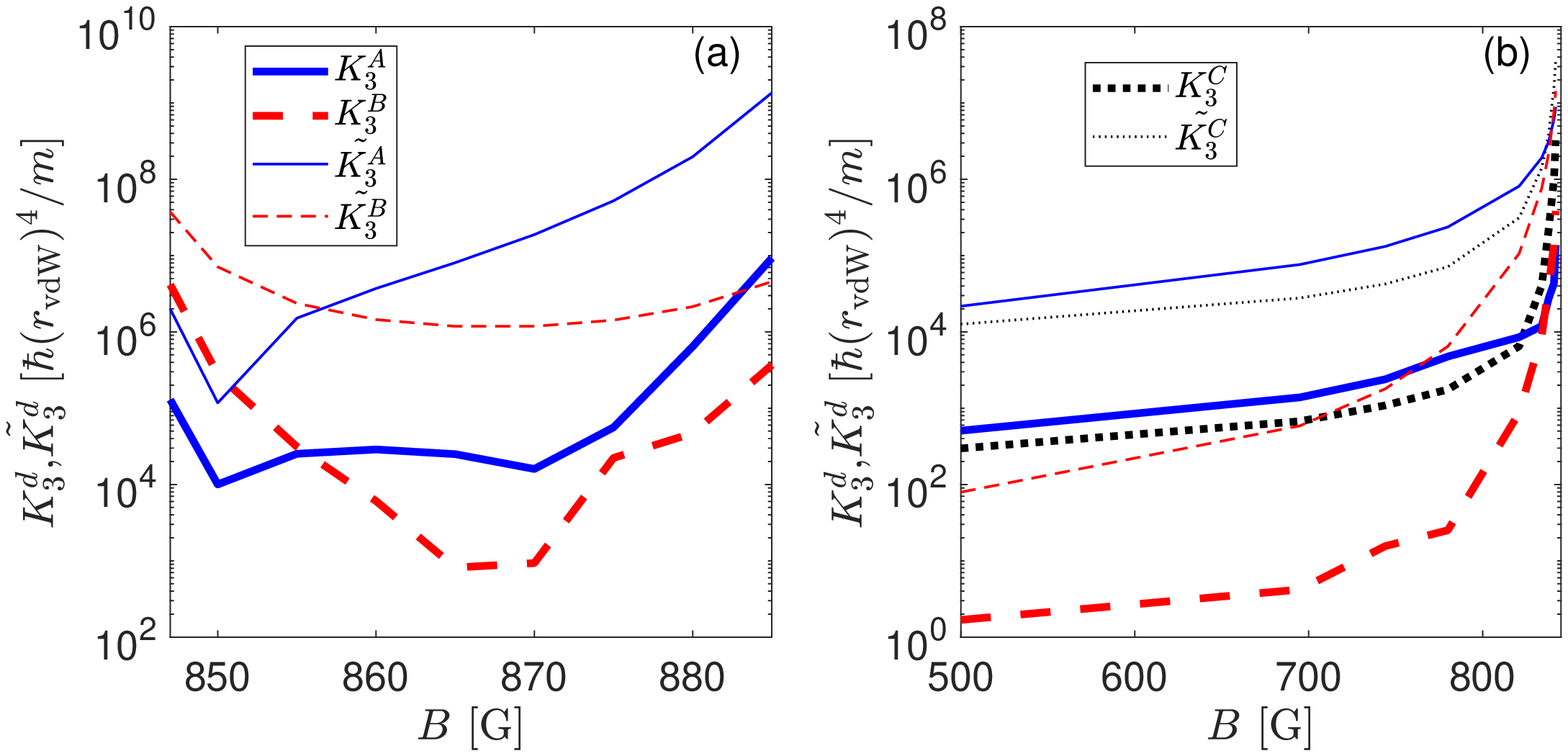} }
\caption{\label{fig:KI} The TBR rates for decay channels A, B and C calculated from $ U_{\alpha 0} $ by using the FMS model with $l_{\rm{max}}=4$ ($K_3^d$) and from $ U^{(1)}_{\alpha 0} $ by using $\mathcal{O}_d$ ($\tilde{K_3^{d}}$).  }
\end{figure*}
To get the expression of $\mathcal{O}_d$ in the main text, we rewrite Eq. (\ref{eq:AGSeqrecom}) as
\begin{equation} \label{eq:Usum}
U_{\alpha 0} (z)=\sum_{n=0}^{\infty}U_{\alpha 0}^{(n)} (z)
\end{equation}
with
\begin{widetext}
\begin{eqnarray} \label{eq:Un}
U_{\alpha 0}^{(n)} (z)= \left\{\left[ P_+ + P_- \right] \mathcal{T}_\alpha (z) G_0(z)\right\}^{n}\frac{1}{3}G_0^{-1}(z) \left[1 + P_+ + P_-\right]. 
\end{eqnarray}
\end{widetext}
Since ${}_\alpha \langle\psi_d | U_{\alpha 0}^{(0)} (z) | \psi_{\rm{in}} \rangle$ vanishes at zero energy, we look into the next order term $U_{\alpha 0}^{(1)} (z)$. The initial free atom state is taken as $|\psi_{\rm{in}}\rangle=|\mathbf{p}=\mathbf{0},\mathbf{q}=\mathbf{0}\rangle|\sigma_a^{\rm{in}}\sigma_b^{\rm{in}}\sigma_c^{\rm{in}}\rangle$, where $\mathbf{p}$ and $\mathbf{q}$ are Jacobi momenta corresponding to the relative motion between two atoms and that of the third atom to the center of mass of them, respectively.  $|\psi_{\rm{in}}\rangle$ is fully symmetric so that
\begin{equation} \label{eq:U1h}
U_{\alpha 0}^{(1)} (z)|\psi_{\rm{in}}\rangle=\left[ P_+ + P_- \right] \mathcal{T}_\alpha (z)|\psi_{\rm{in}}\rangle.
\end{equation}
We implement the partial wave expansion and switch from the plane wave basis $|\mathbf{p},\mathbf{q}\rangle$ to $|p,q\rangle|lLJM_J\rangle$, where $l$ and $L$ are partial wave quantum numbers corresponding to $\mathbf{p}$ and $\mathbf{q}$, respectively. The initial and product states can then be expressed as $|\psi_{\rm{in}}\rangle=\frac{1}{4\pi}|p=0,q=0\rangle |0000\rangle |\sigma_{a}^{\rm{in}}\sigma_{b}^{\rm{in}}\sigma_{c}^{\rm{in}}\rangle$ and $|\psi_d\rangle_\alpha=|\phi_d,q_d\rangle_\alpha |l_dl_d00\rangle|\sigma_{c}^{d}\rangle$, where $\phi_d$ denotes the radial wave function of molecule $d$. We define $|\psi_{\rm{scat}}\rangle_\alpha \equiv |\phi_{\rm{scat}},q=0\rangle_\alpha|0000\rangle|\sigma_c^{\rm{in}}\rangle$ to describe the state of a scattering complex plus a free atom, where $\phi_{\rm{scat}}$ is the radial two-body scattering wave function at zero energy.  Using these states and Eq. (\ref{eq:U1h}), we get
\begin{widetext}
\begin{eqnarray} \label{eq:U1}
{}_\alpha \langle\psi_d | U_{\alpha 0}^{(1)} (0) | \psi_{\rm{in}} \rangle &=&\frac{1}{4\pi}{}_\alpha \langle\psi_d |\left[ P_+ + P_- \right] \mathcal{T}_\alpha (0)|p=0,q=0\rangle |0000\rangle |\sigma_a^{\rm{in}}\sigma_b^{\rm{in}}\sigma_c^{\rm{in}}\rangle \nonumber\\
&=&\frac{1}{4\pi}{}_\alpha \langle\psi_d |\left[ P_+ + P_- \right] V_\alpha |\phi_{\rm{scat}},q=0\rangle |0000\rangle |\sigma_c^{\rm{in}}\rangle \nonumber \\
&=&\frac{1}{4\pi}\mathcal{O}_d
\end{eqnarray}
with
{\allowdisplaybreaks
\begin{eqnarray} \label{eq:Od}
 \mathcal{O}_d&=&{}_\alpha \langle\psi_d |\left[ P_+ + P_- \right] V_\alpha |\psi_{\rm{scat}}\rangle_\alpha  \nonumber\\
 &=&2\sum_{\sigma_{2b}^d,\sigma_{2b}^{\rm{scat}}}{}_\alpha \langle\psi_d |P_+^{\rm{c}}|\sigma_{2b}^{d}\sigma_c^{d}\rangle \langle\sigma_{2b}^{\rm{scat}}\sigma_c^{\rm{in}} |V_\alpha |\psi_{\rm{scat}}\rangle_\alpha  \langle\sigma_{2b}^{d}\sigma_c^{d} | P_+^{\rm{s}}|\sigma_{2b}^{\rm{scat}}\sigma_c^{\rm{in}}\rangle \nonumber\\
 &=&2\sum_{\sigma_{2b}^d,\sigma_{2b}^{\rm{scat}}}\int d \mathbf{q}' \int d \mathbf{q}''{}_\alpha \langle\psi_d |\mathbf{q}''+\frac{1}{2}\mathbf{q}',\mathbf{q}'\rangle |\sigma_{2b}^{d}\sigma_c^{d}\rangle \langle\sigma_{2b}^{\rm{scat}}\sigma_c^{\rm{in}} |\langle-\mathbf{q}'-\frac{1}{2}\mathbf{q}'',\mathbf{q}'' |V_\alpha |\psi_{\rm{scat}}\rangle_\alpha\langle\sigma_{2b}^{d}\sigma_c^{d} | P_+^{\rm{s}}|\sigma_{2b}^{\rm{scat}}\sigma_c^{\rm{in}}\rangle  \nonumber \\
 &=&2\sqrt{2l_d+1}\sum_{\sigma_{2b}^d,\sigma_{2b}^{\rm{scat}}}\langle\sigma_c^d|{}_\alpha \langle\phi_d|\frac{1}{2}q_d,\sigma_{2b}^{d}\sigma_c^{d}\rangle \langle q_d,\sigma_{2b}^{\rm{scat}}\sigma_c^{\rm{in}}|V_\alpha|\phi_{\rm{scat}}\rangle_{\alpha}|\sigma_c^{\rm{in}}\rangle \langle\sigma_{2b}^{d}\sigma_c^{d} | P_+^{\rm{s}}|\sigma_{2b}^{\rm{scat}}\sigma_c^{\rm{in}}\rangle\nonumber \\
  &=&2\sqrt{2l_d+1}\sum_{\sigma_{2b}^d,\sigma_{2b}^{\rm{scat}}}{}_\alpha \langle\phi_d|\frac{1}{2}q_d,\sigma_{2b}^{d}\rangle \langle q_d,\sigma_{2b}^{\rm{scat}}|V_\alpha|\phi_{\rm{scat}}\rangle_{\alpha} \langle\sigma_{2b}^{d}\sigma_c^{d} | P_+^{\rm{s}}|\sigma_{2b}^{\rm{scat}}\sigma_c^{\rm{scat}}\rangle\nonumber \\
 &=&2\sqrt{2l_d+1}\sum_{\sigma_{2b}^d,\sigma_{2b}^{\rm{scat}}}\phi_d^{\sigma_{2b}^d}(\frac{1}{2}q_d)t_{h}^{\sigma_{2b}^{\rm{scat}}}( q_d)\langle\sigma_{2b}^d\sigma_c^d| P_+^{\rm{s}}|\sigma_{2b}^{\rm{scat}}\sigma_c^{\rm{in}}\rangle,
 \end{eqnarray}}
 where
\begin{eqnarray} \label{eq:th}
t_{h}^{\sigma_{2b}^{\rm{scat}}}(q_d)=\langle q_d,\sigma_{2b}^{\rm{scat}}|V_\alpha|\phi_{\rm{scat}}\rangle_\alpha =\langle\sigma_{2b}^{\rm{scat}}|\langle p=q_d|t_\alpha^{l=0}(z_{2b}=0)|p_z=0\rangle|\sigma_a^{\rm{in}}\sigma_b^{\rm{in}}\rangle 
\end{eqnarray}
 \end{widetext}
 is an element of the two-body $s$-wave $t$-matrix $t^{l=0}$ at two-body energy $z_{2b}=p_z^2/m=0$ with one momentum fixed on the energy shell, which is commonly referred to as the half-shell $t$-matrix in nuclear physics \cite{Ernst:1973,Hlophe:2013}. The expression of Eq. (\ref{OD}) in the main text is obtained by filling in $l_d =0$ for $d= A$ or $B$ in Eq. (\ref{eq:Od}). 
 
 Figure \ref{fig:KI}(a) shows that the TBR rates calculated from $\mathcal{O}_d$ follow the overall trend of those given by the FMS calculation with $l_{\rm{max}}=4$ in our considered magnetic field regime. The main feature that three free atoms recombine predominantly into decay channel B at $B \lesssim 855$ G and into decay channel A at $B \gtrsim 855$ G is captured by $\mathcal{O}_d$. Similarly, $\mathcal{O}_d$ captures the overall trend and relative magnitude of the TBR rates near a different zero crossing at $B = 578$ G, as is shown in Fig. \ref{fig:KI}(b). However, the absolute magnitude of the TBR rates cannot be correctly addressed by $\mathcal{O}_d$, indicating that our multichannel numerical calculation is indispensable for quantifying the TBR rates.  Figure \ref{fig:KI}(b) also demonstrates that the spin-exchange recombination pathway to decay channel B is strongly suppressed near the zero crossing at $B = 578$ G. We note that decay channel C in Fig. \ref{fig:KI}(b) corresponds to the new shallow molecule with $M_{2b}=0$ appearing when the magnetic field decreases over the Feshbach resonance position at $B = 845$ G.

 \section{analysis on $\mathcal{C}_{\mathcal{A}_1\mathcal{S}_1}$ and $\mathcal{C}_{\mathcal{B}_3\mathcal{S}_3}$}
  \begin{figure*}[t]
\centering
 \resizebox{0.8\textwidth}{!}{\includegraphics{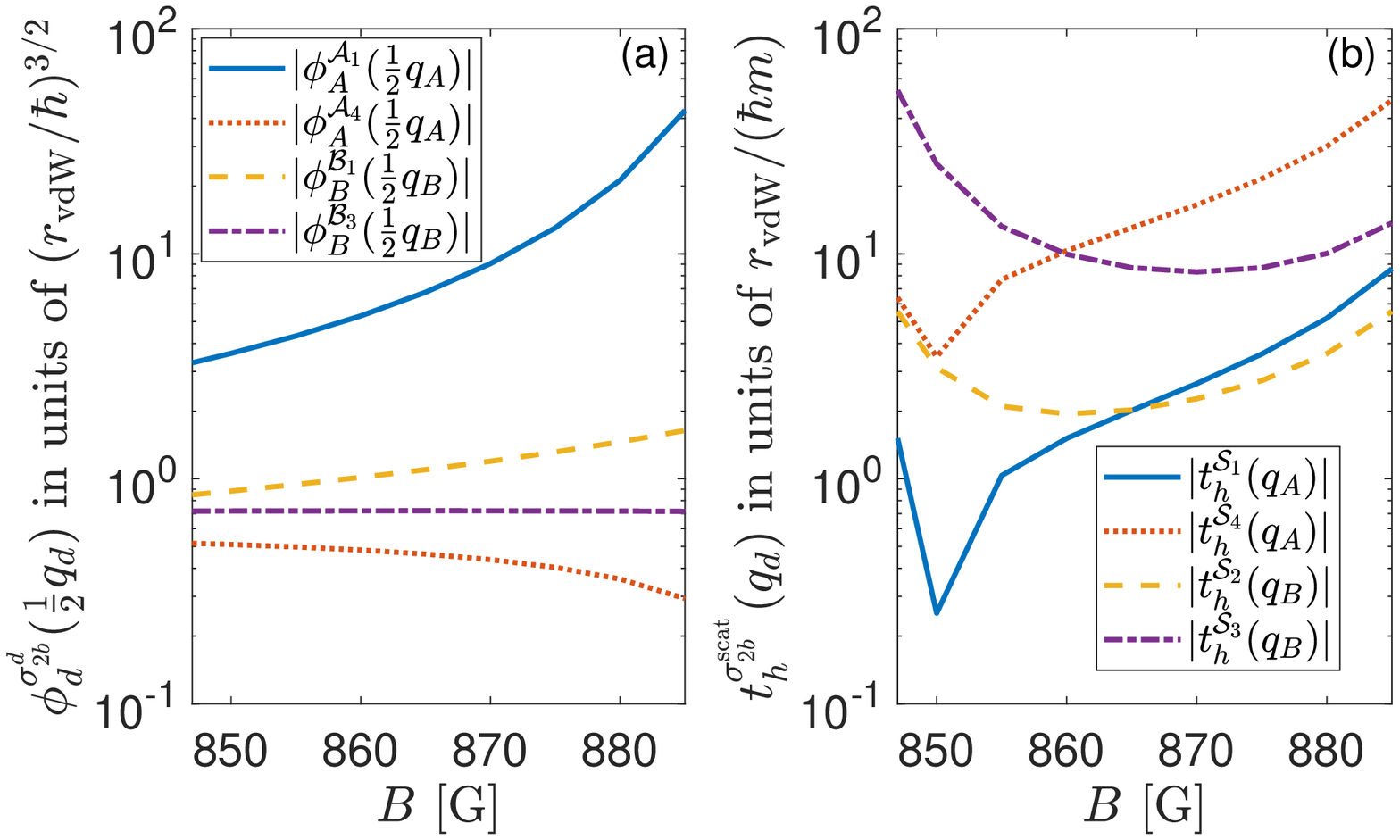} }
\caption{\label{fig:abs} The molecular wave function evaluated at $\frac{1}{2}q_d$ (a) and half-shell $t$-matrix evaluated at $q_d$ (b) relevant for the overlaps $\mathcal{O}_A$ and $\mathcal{O}_B$.}
\end{figure*}

  \begin{figure*}[t]
\centering
 \resizebox{0.9\textwidth}{!}{\includegraphics{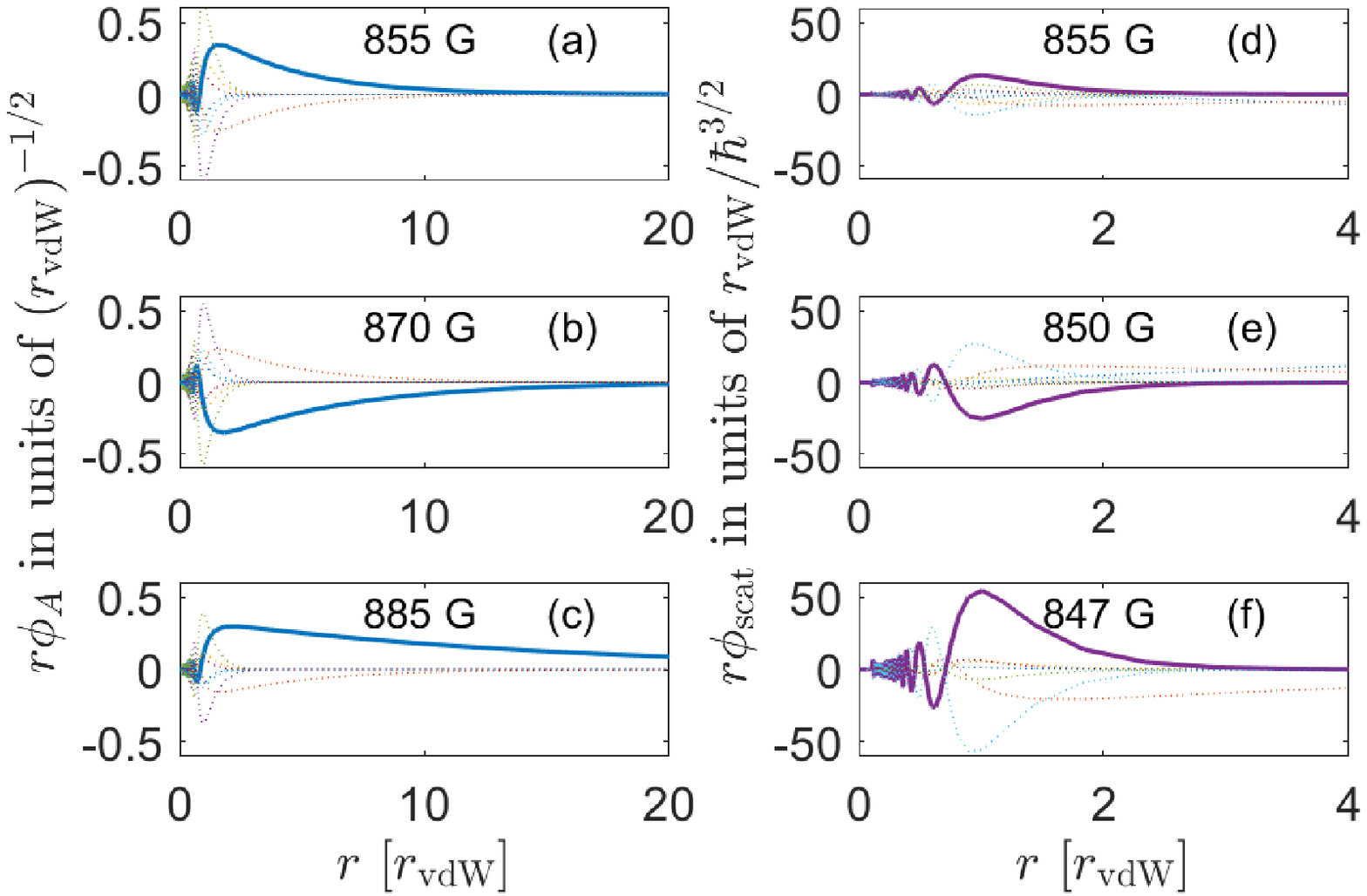} }
\caption{\label{fig:wf} (a)-(c) show the molecular wave function $\phi_A$ at $B= 855, 870$ and 885 G.  (d)-(f) show the two-body scattering wave function $\phi_{\rm{scat}}$ at $B= 855, 850$ and 847 G. The solid lines highlight the $\sigma_{2b}^{\mathcal{A}_1}$ component for $\phi_A$ in (a)-(c) and the $\sigma_{2b}^{\mathcal{B}_3}$ component for $\phi_{\rm{scat}}$ in (d)-(f).}
\end{figure*}
We have demonstrated in the main text that $\mathcal{O}_A$ is determined by $\mathcal{C}_{\mathcal{A}_1\mathcal{S}_1}$ at $B \gtrsim 855$ G and $\mathcal{O}_B$ is determined by $\mathcal{C}_{\mathcal{B}_3\mathcal{S}_3}$ at $B \lesssim 855$ G. Now we want to analyze which quantities make these two components the most significant. For that we write $\mathcal{C}_{\mathcal{A}_1\mathcal{S}_1}$ and $\mathcal{C}_{\mathcal{B}_3\mathcal{S}_3}$ as
\begin{eqnarray}
\mathcal{C}_{\mathcal{A}_1\mathcal{S}_1}&=&\phi_A^{\mathcal{A}_1}(\frac{1}{2}q_A)t_{h}^{\mathcal{S}_1}(q_A), \nonumber \\
\mathcal{C}_{\mathcal{B}_3\mathcal{S}_3}&=&\phi_B^{\mathcal{B}_3}(\frac{1}{2}q_B)t_{h}^{\mathcal{S}_3}(q_A).
\end{eqnarray}
Figure \ref{fig:abs}(a) shows that the large value of $\mathcal{C}_{\mathcal{A}_1\mathcal{S}_1}$ at $B \gtrsim 855$ G comes from $\phi_A^{\mathcal{A}_1}(\frac{1}{2}q_A)$. The increasing behavior of $\phi_A^{\mathcal{A}_1}(\frac{1}{2}q_A)$ with the increase of magnetic field can be understood as follows. Molecule A becomes increasingly extended in the $\mathcal{A}_1$ channel when its energy level is shifted towards the threshold of that channel. Eventually, the energy level of molecule A merges with the threshold of the $\mathcal{A}_1$ channel at the Feshbach resonance position of $B = 894$ G.  As a result, $\phi_A$ increases the amplitude of its $\mathcal{A}_1$ component in the large-distance (or equivalently, low-momentum) regime, as is shown in Figs. \ref{fig:wf}(a)-\ref{fig:wf}(c). In combination with a simultaneously decreasing $q_A$, this leads to a rapid increase of $\phi_A^{\mathcal{A}_1}(\frac{1}{2}q_A)$. 

In contrast, the enhancement of $\mathcal{C}_{\mathcal{B}_3\mathcal{S}_3}$ at $B \lesssim 855$ G with the decreasing magnetic field comes from that of $t_{h}^{\mathcal{S}_3}(q_B)$, as is shown in Fig. \ref{fig:abs}(b). The behavior of $t_{h}^{\mathcal{S}_3}(q_B)$ at $B \lesssim 855$ G can be related to the Feshbach resonance at $B =845$ G. In the vicinity of this Feshbach resonance, the two-body scattering state $|\phi_{\rm{scat}}\rangle$ increases the amplitudes of its closed channel components at short range due to the coupling from the resonant molecular state, as is shown in Figs. \ref{fig:wf}(d)-\ref{fig:wf}(f). This leads to an enhanced component of the two-body half-shell $t$-matrix in the corresponding closed channels. In the present case, $\mathcal{S}_3$ is one of the closed channels with enhanced components. Therefore $t_{h}^{\mathcal{S}_3}(q_B)$ increases when the magnetic field is tuned towards the Feshbach resonance at $B =845$ G. Similarly, $t_{h}^{\mathcal{S}_3}(q_B)$ is also enhanced in the vicinity of the Feshbach resonance at $B = 894$ G.
\end{document}